# Why does pressure melt ice?

**Compression shortens the O:H nonbond and lengthens the H-O bond simultaneously via O-O Coulomb repulsion. The H-O elongation and its energy loss lower the melting point.**

Chang Q Sun, *NOVITAS, School of Electrical and Electronic Engineering, Nanyang Technological University, Singapore 639798*

## Introduction

Regelation, i.e., ice melts under compression and freezes again when the pressure is relieved, remains puzzling since its discovery in 1850's by Faraday, James Thomson and his brother William Thomson (Later Lord Kevin) in 1850's [1,2]. Here we show that hydrogen bond (O:H-O) cooperativity and its extraordinary recoverability resolve this anomaly. The H-O bond and the O:H nonbond possesses each a specific heat $\eta_x(T/\Theta_{Dx})$ whose Debye temperature $\Theta_{Dx}$ is proportional to its characteristic phonon frequency $\omega_x$ according to Einstein's relationship. A superposition of the $\eta_x(T/\Theta_{Dx})$ curves for the H-O bond (x = H, $\omega_H$ ~3200 cm$^{-1}$) and the O:H nonbond (x = L, $\omega_L$ ~ 200 cm$^{-1}$, $\Theta_{DL}$ = 198 K) yields two intersecting temperatures that define the liquid/quasi-solid/solid phase boundaries. Compression shortens the O:H nonbond and stiffens its phonon but does the opposite to the H-O bond through O-O Coulomb repulsion, which closes up the intersection temperatures and hence depress the melting temperature of ice. Reproduction of the $T_m(P)$ profile clarifies that the H-O bond energy $E_H$ determines the $T_m$ with derivative of $E_H$ = 3.97 eV for bulk water and ice. Oxygen atom always finds bonding partners to retain its sp$^3$-orbital hybridization once the O:H breaks, which ensures O:H-O bond recoverability to its original state once the pressure is removed.

Faraday[1] noted that :

'*two pieces of thawing ice, if put together, adhere and become one; at a place where liquefaction was proceeding, congelation suddenly occurs. The effect will take place in air, in water, or in vacuo. It will*

*occur at every point where the two pieces of ice touch; but not with ice below the freezing-point, i.e., with dry ice, or ice so cold as to be everywhere in the solid state'*.

Faraday suggested that there may be a thin liquid-like layer of nascent ice on the surface, ready to be converted to solid on contact with another layer. James Thomson[2] explained this observations in terms of pressure melting based on equilibrium thermodynamics available in his day, and it was his brother, William, who verified the result experimentally[3]. This led to a dispute with Faraday, who observed that blocks of ice would stick together by freezing under mild pressure (which one observes with ice cubes in a basket in modern refrigerators). There is a body of modern literature suggesting that Faraday's surmise of an anomalous ice layer may be correct.

The Regelation can easily be demonstrated by looping a wire around a block of ice with a heavy weight attached to it. This loaded wire melts the local ice gradually until the wire passing through the entire block. The wire's track will refill as soon as it passes, so the ice block will remain solid even after wire passes completely through. Another example is that a glacier can exert a sufficient amount of pressure on its lower surface to lower the melting point of its ice, allowing liquid water flow from the base of a glacier to lower elevations when the temperature of the air is above the freezing point of water. The regelation is exceedingly interesting, because of its relation to glacial action under nature circumstances[4], in its bearing upon molecular action[5], and self-repairing of damaged living cells.

It is usual in 'normal' materials that compression raises the critical temperature ($T_C$) at all phase transitions[6-8]; however, according to the phase diagram of water and ice, the freezing temperature of liquid water is lowered to -22°C by applying 210 MPa pressure; stretching ice (i.e. tensile, or negative, pressure) has the opposite effect - ice melts at +6.5°C when subjected to -95 MPa pressure[9]. Conversely, the $T_C$ for ice drops from 280 to 150 K at the transition from ordered ice-VIII to proton-disordered ice-VII phase when $P$ is increased from 1 to 50 GPa[10-12]. A molecular-dynamics (MD) study of a nanowire cutting through ice suggests that the transition mode and the cutting rate depend on the wetting properties of the wire - hydrophobic and thicker wires cut ice faster[13].

However, a consistent understanding with numerical reproduction of regelation has yet been achieved despite intensive investigations. It might be true that regelation can occur for substances with the property of expanding upon freezing, but mechanisms for both freezing expansion and regelation remain unclear[14]. These issues are beyond the scope of classical thermodynamics in terms of equation of states, which inspires alternative ways of thinking and approaching to unlocking these puzzles.

Recent progress[14-19] enables us to tackle this mystery from the perspective of hydrogen bond (O:H-O) cooperative relaxation under compression. We show in this presentation that the O:H-O bond has extreme recoverability of distortion and dissociation. Numerical reproduction of the pressure dependent melting temperature ($T_m$) of ice revealed that O:H-O bond relaxation disperses the critical temperatures for solid/quasi-solid (traditionally known as liquid-solid transition) phase transition.

**Principle: Hydrogen bond cooperative relaxation**
**General bond potential**

Figure 1a shows a pairing potential u(r) for the interatomic bonding. The coordinates ($d$, $E_b$) at equilibrium are the bond length and bond energy. We are concerned how the $d$ and $E_b$ respond to external stimulus regardless of the shape of the particular u(r). A Taylor series approximates the pairing potential u(r) as follows:

$$u(r) = \left.\frac{\partial^n u(r)}{n!\partial r^n}\right|_{r=d} x^n = E_b + \left.\frac{\partial^2 u(r)}{2\partial r^2}\right|_{r=d} x^2 + \left.\frac{\partial^3 u(r)}{6\partial r^3}\right|_{r=d} x^3 + 0(x^{n\geq 4})$$

(1)

The zeroth differential is the bond energy at equilibrium $E_b$, which can be determined from photoelectron spectrometrics. Higher-order differentials corresponding to the harmonic and nonlinear vibrations determine the shape of the u(r). The vibration amplitude $x$ is 3% or less than atomic distance $d$ of the substance below melting.

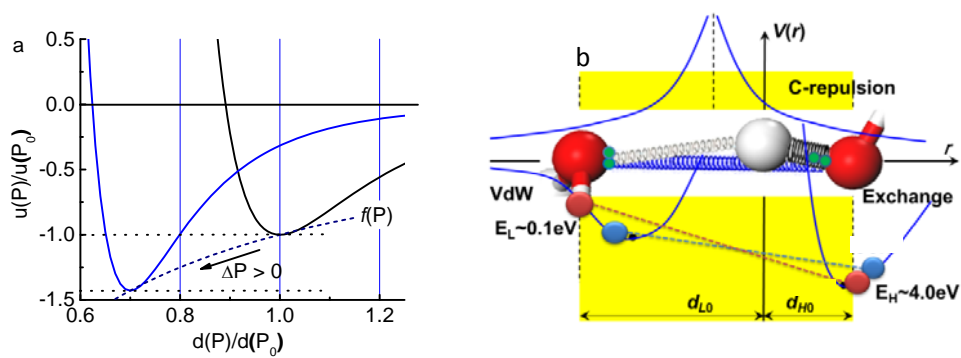

Figure 1. (a) The long-range, mono-well potential for paring atoms in a 'normal' substance and (b) the asymmetric, short-range, double-well potentials for the O:H-O bond and their relaxation dynamics [20,21]. Compression stores energy by shortening and stiffening the bond whereas tension does the opposite, along an $f(P)$ path in (a). O:H-O potentials include the O:H nonbond van der Waals like (vdW-like)

interaction ($E_L \sim 0.1$ eV, left-handed side), the H-O exchange interaction ($E_H \sim 4.0$ eV, right-handed side), and the Coulomb repulsion (C-repulsion) between electron pairs (paring green dots) on oxygen ions. A combination of these interactions with external stimulus dislocates O atoms in the same direction by different amounts. The relaxation proceeds along the potential paths with respect to the H atom (in grey) coordination origin under compression (linked blue spheres) or tension (linked red spheres further moves left). Springs are analogous the respective interactions. The $d_{H0}$ and $d_{L0}$ in (b) are the respective segmental length references at 4 °C.

Generally, external stimuli, such as stressing and heating modulate the length $d(T, P)$ and energy $E(T, P)$ of the representative bond along a path denoted $f(T, P)$ [6]. For instance, compression stores energy into a substance by shortening and stiffening all bonds with possible plastic deformation while tension does the opposite, as illustrated in Figure 1a, and formulated as follows [15]:

$$\begin{cases} d(P,T) = d_b \left(1 + \int_{T_0}^{T} \alpha(t) dt \right) \left(1 - \int_{P_0}^{P} \beta(p) dp \right) \\ E(P,T) = E_b \left(1 - \dfrac{\int_{T_0}^{T} \eta(t) dt + \int_{V_0}^{V} p(v) dv}{E_b} \right) \end{cases}$$

(2)

where $T_0$ and $P_0$ are the ambient referential conditions. The α(t) is the thermal expansion coefficient. $\beta = -\partial v / (v \partial p)$ is the compressibility ($p < 0$, compressive stress) or extensibility ($p > 0$ tensile stress). The $v$ is the volume of a bond (cross sectional area times length). The η(t) is the specific heat of the representative bond in Debye approximation. The integration of the η(t) from 0 K to the melting point ($T_m$) approximates the bond energy by omitting experimental conditions as the η(t) for constant volume deviates only 3% from that of constant pressure [15].

**O:H-O bond asymmetric and short-range potentials**

An extended tetrahedron containing two water molecules and four identical O:H-O bonds has unified the length scale and mass density of molecular packed tetrahedrally in water ice on statistic average [22]. This extension has also turned out the O:H-O bond with asymmetric, short-range O:H, H-O and O---O interactions, see Figure 1b [23]. The O:H-O bond is segmented into a shorter H-O polar-covalent bond with a stronger exchange interaction $u_H(r)$ and a longer O:H nonbond with a weaker nonbond interaction $u_L(r)$. The two segments are coupled by Couomb repulsion between electron pairs on adjacent oxygen atoms

$u_C(r)$ [18,24]. All interactions are limited to the specific segment without any decay acrossing the respective region. The O:H-O bond links the O---O in both the solid and liquid $H_2O$ phase, regardless of phase structures or topologic configurations [22].

The O:H-O bond performs as an asymmetrical oscillator pair. Under the O---O Coulomb coupling, external excitation such as cooling [14], compressing [18], salting [25], and clustering [24] always relaxes the O:H and H-O in the same direction but by different amounts. Because of the strength disparity between the two segments, compression shortens and stiffens the O:H nonbond (left hand side of the O:H-O bond) and simultaneously lengthens and softens the H-O bond (right hand side). The COlomb repulsion makes the O:H-O bond recover completely its initial states once the compression is relieved. Conversely, once the O:H nonbond breaks, oxygen atom finds immediately bonding partner to retain its $sp^3$-orbital hybridization that occurs at temperatures at 5 K [26] and above even in gaseous phase [27].

With the aid of quantum calculations, Lagrangian oscillating mechanics and Fourier fluid thermo dynamics, and phonon spectrometrics, we have been able to consistently and quantitatively resolve quantitatively a few issues such as: 1) Mpemba effect – hot water freezes quicker than its cold [16], 2) supersolid skins for the slipperiness of ice and the hydrophobic and tough skin of water liquid [28], 3) ice expansion and mass density oscillation over full temperatures range [14], 4) anomalies of water molecules with fewer than four nearest neighbors in clusters and droplets[24], 5) Hofmeister effect – NaCl mediation of O-O repulsion [25], 6) density-geometry-dimension correlation of molecules packed in water and ice [22], 7) low compressibility and proton centralization of ice,[18] and, 7) mapping the local potential paths for the O:H-O bond relaxing with stimulus [23], etc. Progress made insofar has formed the subject of a recent treatise [17].

**Results and discussion**

O:H-O bond extraordinary recoverability

Figure 2a shows that a molecular dynamics (MD) decomposition of the measured V-P profile of Ice-VIII at 80 K [29] truns out that the $d_x$ asymmetric relaxation proceeds until proton symmetrization occuring at 0.22 nm and 60 GPa. The subscript x = H and L reresnet for the H-O and the O:H，respectively. The $d_L$ shortens monotonically by 4.3% from 0.1768 to 0.1692 nm and the $d_H$ lengthens by 2.8% from 0.0975 to 0.1003 nm when the pressure is increased from 0 to 20 GPa [18]. The $d_L$ equals the $d_H$ at 0.11 nm and 60 GPa, towards proton centrolization in the O:H-O bond [30-32]. Figure 2b shows the $\omega_x$ cooperative shift of ice under compression at 80 K. Phonon frequencies relax monotonically up to 60 GPa even though the pressure is increased [32,33]. In accordance to the length relxation, compression shifts the $\omega_H$ toward higher frequencies and the $\omega_L$ to lower. The length and stiffness trend of O:H-O bond relxation hold for all phases of water and ice with negligible slope variation [17].

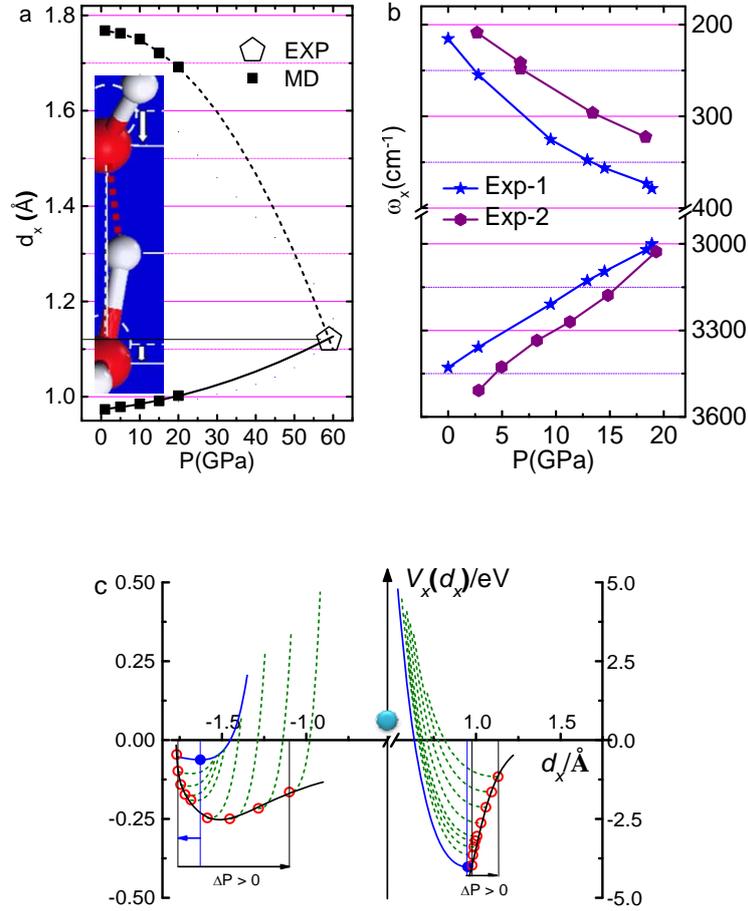

Figure 2. Pressure induced O:H-O bond relaxation in the (a) segmental length $d_x$, (b) phonon frequencies $\omega_x$ [32,33], and (c) potential paths $u_x(r)$ for the O:H-O bond relaxing with pressure (l. to r.: P = 0, 5, 10, 15, 20, 30, 40, 50, 60 GPa)[23]; blue dots correspond to OH-O bond without Coulomb repulsion being involved. The $d_x$ curves in (a) meet at the point of proton centralization occurring in phase X at 59 GPa and 0.22 nm [31,32]. The O:H nonbond and H-O bond responses to compression oppositely (see inset a).

A Lagrangian-Laplace transformation of the measured $d_x$ and $\omega_x$ turns out the force constant $k_x$ and segmental energy $E_x$, which maps the potential paths of the O:H-O bond under compression[23]. As shown in Table 1, compression increase the $E_L$ from 0.046 to 0.250 eV up to 40 GPa and then decrease to 0.16 eV at 60 GPa; the $E_H$ decreases monotonically from 3.97 eV to 1.16 eV at 60 GPa. Different from situation of 'normal' substance, compression lowers the total energy of the O:H-O bond rather than raise it. The O:H-O bond will fully recover its initial states once the compression is relieved without any plastic deformation.

Table 1. Pressure-dependence of the O:H-O segmental cohesive energy $E_x$ and the net gain at each quasi-equilibrium state under compression. Unlike 'normal' substance that gains energy with possible plastic deformation under compression, O:H-O bond always losses energy and tends to recover from its higher energy state to lower initial state without any plastic deformation.

| P (GPa) | $E_L$ (eV) | $E_H$ (eV) | $E_{H+L}(P)-E_{H+L}(0)$ |
|---|---|---|---|
| 0 | 0.046 | 3.97 | 0 |
| 5 | 0.098 | 3.64 | -0.278 |
| 10 | 0.141 | 3.39 | -0.485 |
| 15 | 0.173 | 3.19 | -0.653 |
| 20 | 0.190 | 3.04 | -0.786 |
| 30 | 0.247 | 2.63 | -1.139 |
| 40 | 0.250 | 2.13 | -1.636 |
| 50 | 0.216 | 1.65 | -2.15 |
| 60 | 0.160 | 1.16 | -2.696 |

As expected, compression shortens the $d_L$, increases the $\omega_L$ and $E_L$ of the O:H nonbond; the H-O bond responses oppositely to compression, resulting in $d_H$ elongation, $\omega_H$ and bond energy $E_H$ reduction, which can be formulated in the reduced forms as follows ($E_x$ valids at P < 30 GPa; $d_x/d_{x0} = 1 + \beta_{x1}P + \beta_{x2}P^2$ for instance):

$$\begin{pmatrix} d_H/0.9754 \\ d_L/1.7687 \\ \omega_H/3326.140 \\ \omega_L/237.422 \\ E_H/3.970 \\ E_L/0.046 \end{pmatrix} = 1 + \begin{pmatrix} 9.510 \times 10^{-2} & 0.2893 \\ -3.477 \times 10^{-2} & -1.0280 \\ -0.905 & 1.438 \\ 5.288 & -9.672 \\ -1.784 & 3.124 \\ 25.789 & -49.206 \end{pmatrix} \begin{pmatrix} 10^{-2} P^1 \\ 10^{-4} P^2 \end{pmatrix}$$

(3)

**$E_H$ dictating the $T_m$**

The following proves that $E_H$ dictates the $T_m$ for melting, $T_m \propto E_H$. Acceding to eq (2), The $T_m$ changes in the following relationship but $x = L$ or $H$ is yet to be known [6],

$$\frac{T_C(P)}{T_C(P_0)} = 1 - \frac{E_x(P)}{E_{x0}} = 1 - \frac{\int_{V_0}^{V} p\,dv_x}{E_{x0}} = 1 - \frac{s_H \int_{P_0}^{P} p\frac{dd_H}{dp} dp}{E_{H0}} < 1.$$

(4)

Eq (3) defines the slope of $d_x(P)$:

$$\frac{dd_x}{dp} = d_{x0}[\beta_{x1} + 2\beta_{x2}P].$$

(5)

Generally, pressure raises the $T_m$ but ice responses to pressure in the opposite – $T_m$ drops when the pressure is increased. Reproduction of the measured $P$-dependent $T_m$ for melting (Figure 3a) [34] requires that the integral in eq (4) must be positive. Only the $d_H$ in Eq. (3) meets this criterion ($\beta_{x1} > 0$ and $\beta_{x2} > 0$). Therefore, the H-O bond $E_H$ dominates the $T_m$.

Furthermore, matching the $T_m(P)$ profile using Eq. (5) yields an $E_H$ value of 3.97 eV at 0.1 MPa(1 atm pressure) by taking the H atomic diameter of 0.106 nm as the diameter of the H-O bond [35]. This $E_H$ value agrees with the energy of 4.66 eV for dissociating the H-O bond of water molecules deposited on a TiO$_2$ substrate with less than a monolayer coverage, and 5.10 eV for dissociating water monomers in the gaseous phase [36]. Molecular undercoordination shortens the H-O bond and raises its cohesive energy from the bulk value of 3.97 to 4.66 and to 5.10 eV when the O:H-O bond is subject to molecular undercoordination[25].

Clearly, the relaxation of the H-O bond mediates the $T_m$, while $E_L$ is largely irrelevant. It is not surprising, therefore, that compression softens the H-O bond and hence lowers the $T_m$, while negative (tensile) pressure does the opposite by shortening and stiffening the H-O bond [34], and hence negative pressure elevates the $T_m$.

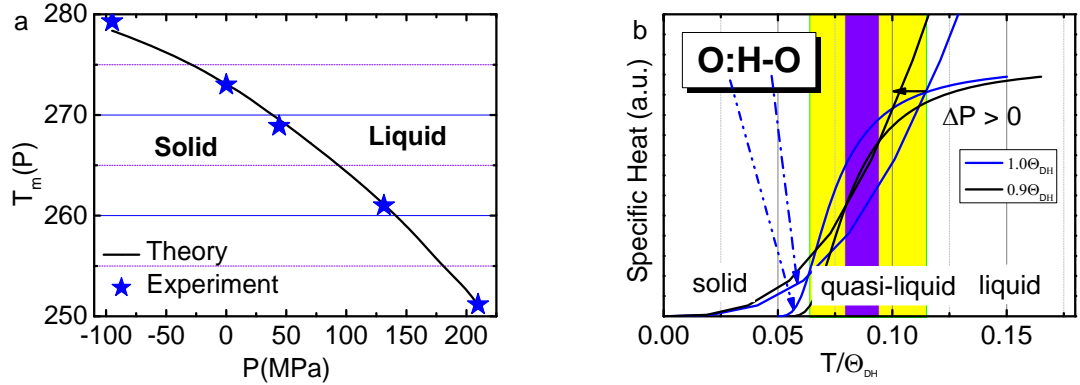

Figure 3. (a) Theoretical reproduction of the measured $T_m(P)$ (-22°C at 210 MPa; +6.5°C at -95 MPa) [34] profiles confirms that the $E_H$ dictates the $T_m$ for ice melting with derivative of $E_H$ = 3.97 eV for bulk water [18]. (b) The superposition of the $\eta_x(T)$ curves yields two crossing temperatures that defines the solid/quasisolid/liquid phase boundaries. The high temperature boundary corresponds to quasisolid melting and the lower to freezing. Compression/tension ($\Delta P > 0$)/($\Delta P < 0$) disperses the boundaries simultaneously and reversely by modulating the $\Theta_{Dx} \propto \omega_{Dx}$ and $E_x \propto \int_0^{T_{mx}} \eta_x(t)dt$, depressing/elevating the $T_m$.

## $T_m(E_H)$ and $T_V(E_L)$ paradox: phase-boundary dispersivity

It is known that evaporating one $H_2O$ molecule from bulk water requires energy of $4E_L$ = 0.38 eV [37] to break four O:H nonbonds surrounding the molecule. This happens at the ambient pressure and $T_V$ = 373 K temperature. Question may arise why the $E_H$ instead of the $E_L$ dominates the $T_m$ though the $T_V$ is higher than the $T_m$?

In order to clarify this paradox, let us look at the specific heat of water [14]. Generally, the specific heat of a 'normal' substance is regarded as a macroscopic quantity integrated over all bonds of the specimen, which is also the amount of energy required to raise the temperature of the substance by 1 K degree.

However, in dealing with the representative for all bonds of the entire specimen, it is necessary to consider the specific heat per bond that is obtained by dividing the bulk specific heat by the total number of bonds [6]. For a specimen of other usual materials, one bond represents all on average; therefore the thermal response is the same for all the bonds, without any discrimination among all bonds in cooling contraction and thermal expansion [38].

For water ice, however, the representative O:H-O bond is composed of two segments with strong disparity in the specific heat of the Debye approximation, $\eta_x(T, \Theta_{Dx})$ [14]. These two segments response to a thermal excitation differently. Two parameters characterize the specific heat curves each. One is the Debye temperature $\Theta_{Dx}$ and the other is the thermal integral of the $\eta_x(T, \Theta_{Dx})$ from 0 K to the $T_{mx}$. The $\Theta_{Dx}$ determines the rate at which the specific-heat curve reaches its saturation. The $\eta_x(T, \Theta_{Dx})$ curve of a segment with a relatively lower $\Theta_{Dx}$ value reaches saturation more rapidly than the other segment, since the $\Theta_{Dx}$, which is lower than $T_{mx}$, is proportional to the characteristic vibration frequency $\omega_x$ of the respective segment, $k\Theta_{Dx} = \hbar\omega_x$, according to Einstein's relation, [39] where $k$ and $\hbar$ are constants.

Conversely, the integral of $\eta_x(T, \Theta_{Dx})$ from 0 K to the $T_{mx}$ determines the cohesive energy per bond $E_x$ [6]. The $T_{mx}$ is the temperature at which the vibration amplitude of an atom or a molecule expands abruptly to more than 3% of its diameter irrespective of the environment or the size of a molecular cluster [39,40]. Thus we have:

$$\begin{cases} \Theta_{DL} / \Theta_{DH} \approx 198/\Theta_{DH} \approx \omega_L / \omega_H \approx 200/3200 \sim 1/16 \\ \left(\int_0^{T_{mH}} \eta_H dt\right) / \left(\int_0^{T_{mL}} \eta_L dt\right) \approx E_H / E_L \approx 4.0/0.1 \sim 40 \end{cases}.$$

(6)

Analysis of the temperature-dependence of water surface tension[37] yielded $\Theta_{DL}$ = 198 K < 273 K ($T_m$) and $E_L$ = 0.095 eV compared with $E_H$ = 3.97 eV for bulk water ice [25]. Hence, $\Theta_{DH} \approx 16 \times \Theta_{DL} \approx 3200$ K. The O:H specific heat $\eta_L$ ends at 273 K and the H-O specific heat $\eta_H$ ends at T ≥ 3200 K ($T_{mH}$). The area covered by the $\eta_H$ curve is 40 times greater that covered by the $\eta_L$ curve.

The superposition of these two $\eta_x(T, \Theta_{Dx})$ curves implies that the heat capacity of water ice differs from that of other, 'normal', materials. Such a $\eta_x(T, \Theta_{Dx})$ disparity yields temperature regions with different $\eta_L/\eta_H$ ratios over the full temperature range; see Figure 3b. These regions correspond to phases of liquid and solid ($\eta_L/\eta_H$ < 1), and quasisolid ($\eta_L/\eta_H$ > 1). The intersecting temperatures ($\eta_L/\eta_H$ = 1) correspond to extreme densities at boundaries of the quasisolid phase (viscose and jelly like). The high-temperature boundary corresponds to the maximal density at 4 °C and the lower to the crystallization of bulk water.

Numerical and experimental observations [14,17,22] confirmed that cooling shortens the O:H nonbond in the liquid phase at temperature above 4 °C and in the solid phase below 258 K for bulk at different rates because $\eta_L/\eta_H < 1$ in both regime. However, Cooling shortens the H-O bond in the quasisolid phase (277-258 K). The other counterpart in the O:H-O bond responses to cooling in the opposite direction. This observation clarifies that the segment with lower $\eta_x$ value follows the general rule of thermal expansion and drives the thermal relaxation of the O:H-O bond, which evidences the essentiality of considering the disparity of the specific heat of water ice [14].

One can imagine what will happen to the crossing temperatures if one depresses the $\Theta_{DH}(\omega_H)$ and $E_H$, and meanwhile, elevates the $\Theta_{DL}(\omega_L)$ and $E_L$ by compression or the inverse. Compression ($\Delta P > 0$) raises the $\Theta_{DL}$ and $E_L$ by stiffening $\omega_L$, and meanwhile, lowers the $\Theta_{DH}$ and $E_H$ by stiffening $\omega_L$; however, tension ($\Delta P < 0$) does the opposite. Figure 3b illustrates how the positive P squeezes the quasisolid phase boundaries. The $E_H$ determines approximately the $T_m$ through dispersing the upper phase boundary. The $\Theta_{Dx}(\omega_x)$ always relax simultaneously in opposite direction under a given stimulus, which will disperse the quasisolid phase boundaries resulting in the observed 'superheating/supercooling', as one often refers. In fact, external stimulus can raise/depress the melting/freezing point by phonon relaxation, which is different from the effect of superheating/supercooling [41].

Once the O:H bond breaks, oxygen atoms will find new partners to retain the sp$^3$-orbital hybridization, which is the same to diamond oxidation and metal corrosion – oxygen atoms penetrate into the bulk when corrosion occurs [15,27]. Therefore, O:H-O bond has the strong recoverability for O:H-O bond relaxation and dissociation without any plastic deformation.

## Conclusion

Numerical reproduction of the pressure effect on $T_m$ clarifies that O:H-O bond relaxation in length, energy, and phonon frequency disperses the quasisolid phase boundaries defined by the supposition of the $\eta_x(T)$ curves. Compression stiffens the O:H nonbond and softens the H-O bond, which closes up the separation between the crossing points and depresses the melting temperature of ice. Negative pressure does the opposite to raise the $T_m$. Numerical duplication of the $T_m(P)$ gives rise to the H-O bond cohesive energy of 3.97 eV for the bulk water and ice. Unlike 'normal' substance that gains energy with potential plastic deformation under compression, O:H-O bond demonstrates extreme recoverability of relaxation and dissociation because of not only the nature of oxygen $sp^3$-orbital hybridization but also energy loss at compressed state. The O:H-O always tends to recover from its higher-energy state to initially lower state. Coulomb repulsion between electron pairs on adjacent oxygen ions and the O:H-O bond segmental disparity form the soul dictating its adaptivity, cooperativity, sensitivity, memory, and recoverability when subject to stimulus. Observations may extend to damage recovery of living cells of which O:H-O bond dominates.